\documentclass[preprint]{revtex4}
\usepackage{epsfig,latexsym,amssymb}
\usepackage{latexsym}
\usepackage{amssymb}
\usepackage{amsfonts}
\usepackage{comment}
\usepackage{graphicx}
\usepackage{verbatim}

\newcommand{\be}{\begin{equation}}
\newcommand{\ee}{\end{equation}}
\newcommand{\bea}{\begin{eqnarray}}
\newcommand{\eea}{\end{eqnarray}}

\begin{document}

\title{Axions/Axion like particles and the CMB asymmetric dipole}

\author{Qiaoli Yang}
\email{qiaoliyang@phys.ufl.edu}
\affiliation{Department of Physics, Huazhong Univ. of Science and Technology, Wuhan, 430074}

\begin{abstract}
Axions and axion like particles are very attractive dark matter candidates. In this review, we briefly investigate how the cosmological observations reveal the existence of dark matter and some unique properties of axions/axion like particles which make them more interesting.
\end{abstract}

\maketitle

\section{Cosmology reveals dark matter}
Modern cosmology was born upon the discovery of the General Relativity. Observations reveal that our universe was once much hotter and denser. The Einstein equation describes the relationship between the space-time geometry and the energy density which is written as:
\begin{equation}
G^\alpha_\beta=8\pi GT^\alpha _\beta
\end{equation}
in which $
G^\alpha _\beta=R^\alpha _\beta-{1\over2}\delta^\alpha _\beta R-\Lambda\delta^\alpha _\beta$
and $\Lambda$ is the cosmological constant. Cosmology principle determines the unperturbed metric tensor as:
\begin{equation}
 ds^2=-dt^2+a^2(t)({dr^2\over{1-kr^2}}+r^2d\theta^2+r^2sin^2\theta d\phi^2),
 \end{equation}
in which $a(t)$ is the scale factor and $k$ is a real number that determines the curvature. If contents of our universe can be described as a perfect fluid, the stress-energy tensor can be written as: $T^{\alpha}_{\beta}=(\rho+p)\delta^{\alpha}_{\beta}+p\eta^{\alpha}_{\beta}$,
in which $\rho$ and $p$ are the energy density and the pressure respectively. From equations above one gets two independent equations which are the Friedmann's equation:
\begin{equation}
{3\over a^2}(\dot a^2+k)=8\pi G\rho+\Lambda~~,
\label{eq0}
\end{equation}
and
\begin{equation}
{\ddot a\over a}=-{4\pi G\over3}(\rho+3p)+{\Lambda \over 3}~~.
\label{eq1}
\end{equation}
By combining the two equations above and the relations of the pressure and the energy density, we have the dependence of the energy density upon the scale factor. For dust like matter $p=0$ therefore $\rho_m(t)=\rho_{m,0}/a^3$ and for radiation $p=\rho/3$ which means $
\rho_r(t)=\rho_{r,0}/a^4.$ Thus Eq.(\ref{eq0}) can be rewritten as:
\begin{equation}
({\dot a\over a})^2={8\over3}\pi G ({\rho_{r,0}\over a^4}+{\rho_{m,0}\over a^3})-{k\over a^2}+{\Lambda\over 3}~~,
\end{equation}
from which we find correspondences between scale factors and dominate eras. When $a(t)$ increasing firstly the radiation dominates, then the matter dominates, then the space-curvature term $k$ dominates, and finally the cosmological constant dominates.

Rewriting the Friedmann's equation we have:
\begin{eqnarray}
k=({8\pi G\over3}\rho+{\Lambda\over3}-H^2)a^2={8\pi G a^2\over 3}(\rho+{\Lambda\over8\pi G}-\rho_c)
\end{eqnarray}
where $H=\dot a(t) /a(t)$ is the Hubble parameter and $\rho_c=3H^2/8\pi G$ is the critical density. Now we can define the total energy density: $\rho_t=\rho+{\Lambda\over8\pi G}$ and the ratio of the total energy density to the critical density $\Omega=\rho_t/\rho_c$. We see that the curvature of geometry and the density ratio are related by simple relations:
if $\Omega>1\to k>0$, if $\Omega=1\to k=0$ and if $\Omega<1\to k<0$. So we can draw an interesting conclusion that the ratio of energy density $\Omega$ determines the curvature of our universe. Also we can identify three types of energy density contributed to $\Omega$: 1) radiation $\rho_r(t)$; 2) matter $\rho_m(t)$; and 3) dark energy $\rho_d={\Lambda\over8\pi G}\sim~{\rm constant}$. Observations suggested that $\Omega_{t}=1.0023^{+0.0056}_{-0.0054}$ so our universe is flat which is consistent with the inflation paradigm. There are at least three types of eras our universe can experience: 1) The radiation dominated era in which the scale factor $a(t)_{r}\propto t^{1/2}$; 2) The matter dominated era in which the scale factor $a(t)_{m}\propto t^{2/3}$; and 3) The dark energy dominated era in which the scale factor $a(t)_{\Lambda}\propto e^t$.

Photons propagate on-shell therefore we have: ${dt\over a(t)}=-{dr\over \sqrt{1-kr^2}}$ for FRW matrix and ${\nu_0\over \nu}={a(t)\over a(t_0)}$.
In literature, people denote $z= {\nu_0\over \nu}-1$ to define the redshift, so one finds: $1+z={a(t_0)\over a(t)}$. At the same time, we have: $d_L=\sqrt{L\over4\pi B}$ where $L$ is the emitting power of source, and $B$ is the observed apparent luminosity. Considering a source emitting $N$ photons with a frequency $\nu$ during time duration of $dt$, we have: $L={Nh\nu\over dt}$. So the observed luminosity at time $t_0$ is: $
B={Nh\nu\over 4\pi a(t_0)^2r^2 dt_0}$. Thus we have:
$
d_L=a(t_0)r(1+z)$,
which relates the luminosity distance $d_L$ and the cosmological redshift $z$. Now let us consider the relation between $r$ and $z$. We have: $dt=da/\dot a$, $a(t_0)=1$ and $a(t_1)=(1+z)^{-1}$, which imply
$-\int^0_r{dr\over \sqrt{1-kr^2}}=\int^{t_0}_t{dt\over a(t)}={1\over a(t_0)}\int^1_{(1+z)^{-1}}{da\over a\dot a}$.
Combing the Friedmann equation, relations between $d_L$ and $z$, $r$ and $z$, and the fact that $\Omega_{r0}<<1$, we have:
\begin{equation}
H_0d_L={1+z\over |\Omega_k|^{1/2}}sinn\{ |\Omega_k|^2\int^z_0{dx\over (1+x^2)(1+\Omega_{m0}x)-x(2+x)\Omega_\Lambda} \}
\label{eqH0}
\end{equation}
where $H_0=\dot a(t_0)/a(t_0)$ is the Hubble constant, $\Omega_k=1-\Omega_{m0}-\Omega_{\Lambda}$, and
$sinn=1$, if $k=0$.
When $z$ is small, we can expand Eq.(\ref{eqH0}) as:
\begin{equation}
H_0d_L=z+{1\over2}(1+\Omega_\Lambda-\Omega_{m0}/2)z^3+...~~,
\end{equation}
in which we find that the first term of RHS is the Hubble's law. For large redshift $z$ one can fit the relationship of the luminosity distance and the redshift $z$ to determine $\Omega_\Lambda$ and $\Omega_{m0}$. In the late 90's Perlmutter, Schmidt, and Riess et al. fitted the luminosity distance-redshift function of type Ia supernovae with a very high accuracy and found a dark energy dominated universe where $\Omega_\Lambda\sim74{\%}$. Thus we conclude that the matter takes $26\%$ of total energy density.

Now let us consider the constitutes within the $26\%$. The condition of particles in a thermal equilibrium is: $\Gamma>>{1\over \Delta t}\sim H$,
since thermalization means the particle collision rate is larger than the expansion rate of universe. The collision rate of particles can be calculated:
\begin{equation}
\Gamma=n<v\sigma(v)>
\end{equation}
where $n$ is the particle number density, $v$ is particle's relative velocity, $\sigma(v)$ is the cross-section and $<>$ denotes the average over velocities. Therefore the decoupling temperatures are determined by the couplings of particles. Let us consider a simply example: the neutrinos are coupling to leptons and baryons with a cross-section of order $\sigma\sim G_F^2T^2$, where $G_F$ is the Fermi coupling. The velocity of neutrinos is ultra-relativistic $v\sim1$ and their number density is $n\sim T^3$. So we can calculate the collision rate as: $\Gamma\sim G_F^2T^5$. Since the Hubble rate is $H=\dot a/a\sim1/t\sim G^{1/2}T^2$, we have:
$
{\Gamma\over H}\sim G_F^2G^{-1/2}T^3~~.
$
We can see that the decoupling temperature of neutrinos is $T\sim G_F^{-2/3}G^{1/6}\sim1{\rm MeV}$.

For ultra-relativistic particles such as photons, their energy density is:
$
\rho=g\int{d^3p\over (2\pi)^3}{p\over e^{p/T}\bar+1}
$
in which $g$ is the number of spin degrees of freedom, $T$ is their temperature, and $-$ for bosons, $+$ for fermions. So we have:
$
\rho_b={\pi^2\over30}g_bT^4~~{\rm for~bosons}$ and
$\rho_f={\pi^2\over30}{7\over8}g_fT^4~~{\rm for~~fermions}.
$ Our cosmic photons are hotter than the cosmic neutrinos because annihilation of electrons and positrons injected energy into the photons when the neutrinos had already decoupled. One can calculate the temperature ratio between photons and neutrinos by means of entropy conservation: $
{T_\nu\over T_\gamma}=({g_1\over g_0})^{1/3}=({4\over11})^{1/3}
$
where $g_0$ and $g_1$ are the effective number of degrees of freedom before and after electron-positron annihilation. Therefore the total radiation energy density of radiation is:
\begin{equation}
\rho_r={\pi^2\over15}\cdot(1+3.046{7\over8}({4\over11})^{4/3})T^4~~.
\label{nev1}
\end{equation}
The photon temperature today is $T=2.73{\rm K}$, so the energy density of radiation today is $\rho_r\sim4.7*10^{-34}{\rm g/cm^3}$. Since the critical density is $\rho_c={3H_0^2\over 8\pi G}\sim2*10^{-29}{\rm g/cm^3}$, the ratio of radiation energy density to critical density is of order $10^{-5}$ which is negligible. The Eq.(\ref{nev1}), however, is not completely applicable today since the neutrinos are massive with masses bigger than their current kinetic energy $\sim {\rm K}$. So the energy density of neutrinos today is $n*(\Sigma m_n)$ instead where $n$ is the number density of the neutrinos and $m_n$ are neutrino masses for each species respectively. The number density ratio between neutrinos and photons is unchanged after the electron-positron annihilation so we have the energy density of neutrinos today: $\rho_n=(3/22)n_{\gamma}*(\Sigma m_n)$ where $n_\gamma$ is the photon number density. With current upper limit on neutrinos masses, this contribution to the total energy density is still subdominant.

Let us consider the abundant of baryonic matter. The primordial nucleosynthesis began after the cosmological temperature dropped to $0.05$MeV with typical reactions such as:
$
n+ \nu \leftrightarrow p+e,~
n+e \leftrightarrow p+\bar\nu,$
to produce neutrons, and
$
p+n\leftrightarrow D+\gamma ,$
$D+D\to {^3He}+n,$
$^3He+n\to {^3T}+p,
...$
The baryon number became stabilized after baryon anti-baryon annihilated in the early universe and the photon number became a constant after the decoupling. The ratio of baryon number to the number of photons is $\eta=n_B/n_{\gamma}$. In literature it is convenient to use $\eta_{10}=10^{10}\eta$ since $\eta$ is a very small number. So from the photon number $n_{\gamma}\sim T^3$, one can determine the baryon number if $\eta$ is known. The collisions between nucleons form new heavier nucleons if the new ones were not dissociated by surrounding high energy photons. The processes are mathematically described as:
\begin{equation}
dn_i/dt=-3Hn_i+\sum_{a,j}n_an_j<v\sigma_{aj\to i}>,
\end{equation}
where the $n_i$ are the number densities of respective particles. The equations can apply after the deuterons were not dissociated by photons which depends on $\eta_{10}$. Spergel. et al. used this technology to find $\eta_{10}=6.3\pm0.3$ and the predicted abundances of light elements are consistent with direct observations. All these consistent observations and calculations give us a high confidence on the baryonic energy density which is about $4\%$ of critical density.

Now we have a very important question: what are the majority constituents of matter in the universe? We know that part of matter is baryonic, however their energy density is order of $4\%\rho_c$ as suggested from the big bang nucleosynthesis. The energy density of radiations are too small to be considered. So there are a lot particles in our universe \cite{2014A&A...571A..16P} which is not a part of the standard model.

\section{Axions and axion like particles can be the dark matter}
The axions \cite{Weinberg:1977ma, Wilczek:1977pj,Kim:1979if, Shifman:1979if,Zhitnitsky:1980tq,Dine:1981rt} were proposed from the Peccei-Quinn mechanism which was intended to solve the strong CP problem \cite{Peccei:1977hh}. The Peccei-Quinn mechanism introduces a new $U(1)$ symmetry and the axions are the Goldstone bosons created after the breaking of the new $U(1)$ symmetry with a small mass due to the QCD instanton effects. Axion like particles (ALPs) \cite{Svrcek:2006yi} are created from compactified antisymmetric tensor fields on closed cycles in the string theory. They are the zero Kaluza-Klein mode of the compactified field but acquire small masses due to non-perturbative effects on the cycles \cite{witten:1986,Becker:1995kb,Kallosh:1995hi}. Properties of QCD Axions and ALPs are alike but ALPs have a much less constrained parameter space \cite{Kaloper:2008qs}. The axions and/or ALPs created from the re-alignment mechanism can be a substantial fraction of dark matter \cite{Ipser:1983mw, Preskill:1982cy, Abbott:1982af, Dine:1982ah, Berezhiani:1990sy, Sikivie:2006ni,Mazumdar:2015pta,Yang:2015qka}

The axion phenomenologies are determined by a crucial factor, the symmetry breaking scale $f_a$. Observational and experimental constraints on the $f_a$ is $10^9{\rm GeV}<f_a<10^{12}{\rm GeV}$ \cite{Dicus:1978fp,Vysotsky:1978dc,Dicus:1979ch,Raffelt:1987yu,Dearborn:1985gp,Khlopov:1999tm}.
The following Lagrangian density can be used for the studies of axion cold dark matter:
\be
\mathcal{L}={1\over2}(\partial a)^2-{1\over 2}m^2a^2-{\lambda\over 4!}a^4~~.
\ee
The cold axions were created by the mis-alignment if the inflation happened after the PQ symmetry breaking so all topological defecates were blown out
without a contribution. The respective abundance is of order \cite{Sikivie:2006ni}:
$
\Omega_a\sim({f_a \over {10^{12} {\rm GeV}}})^{7/6}
$.
The effective Lagrangian for ALPs in four dimension is similar to the axions which is:
\be
{\cal L}={f^2_{ALPs}\over 2}(\partial a)^2-\Lambda_{ALPs}^4U(a)~~.
\ee
The $f_{ALPs}$ is the decay constant and the $\Lambda_{ALPs}$ is the potential energy scale for ALPs. Due to the potential energy scale is depending on UV energy scales of string theory and exponentially on string instontons so their mass ranges are scattered to a large energy scale which is one of the major phenomenological difference between axions and ALPs.

\section{Evolutions of axions/ALPs cold dark matter}
Cosmic axions/ALPs are highly occupied in the phase space so we can use the classical field approximation to study evolutions of these particles before thermalization processes happened. In the flat FRW universe the Lagrangian of axions or ALPs is:
\be
{\cal L}={1\over2}\dot a^2R^3-{1\over 2}\partial _i^2 a R-{1\over 2}m^2a^2R^3-{\lambda\over 4!}a^4R^3
\ee
in which $R$ is the scale factor. So the equation of motion is:
$
\partial^2_ta-{1\over R^2}\nabla^2 a+3H\partial_ta+m^2a+{\lambda\over6}a^3=0~~.
$
We are interesting in the slow varying terms of the field since cold dark matter is non-relativistic so the rapid varying terms of order of particle masses can be factored out, thus we have:
\be
-i\dot \psi-{1\over 2m R^2}\nabla^2\psi-i{3\over 2}H\psi+{\lambda\over 8m^2}|\psi|^2\psi=0~.
\label{waveeq}
\ee
This wave equation is very similar to a non-linear schrodinger equation but instead of a description of a single particle probability it determines evolutions of a classical wave. We can use classical analogues of the density and the velocity to study the wave. Thus from eq.(\ref{waveeq}) we found the equation for the density:
\be
{\partial n_a\over \partial t}+{1\over R}{\partial(n_a v^i)\over \partial x^i}+3Hn_a=0
\label{density}
\ee
and the equation for the first order velocity:
\be
{\partial v^i\over \partial t}+Hv^i+{\lambda \over 8m^3}\partial_i n_a-{1\over 2m^2}\partial_i{{\partial_i}^2\sqrt n_a\over \sqrt n_a}=0
\label{velocity}
\ee
The equation of density is the same as that of point like CDM particles but the first order velocity equation has two additional terms ${\lambda \over 8m^3}\partial_i n_a$ and $-{1\over 2m^2}\partial_i{{\partial_i}^2\sqrt n_a\over \sqrt n_a}$ compared with the point like particles. These two additional terms come from the particles self interaction and the quantum pressure respectively. For QCD axions these two terms are too small for observable scales however for ALPs they can have important observable consequences \cite{Yang:2015paa}.

\section{Axions/ALPs may explain the anomalies of CMBR}
The cosmic axions/ALPs have unique properties that they are bosonic and are highly occupied in the phase space. Therefore they can form a Bose-Einstein condensate if they thermalize during the evolution of our universe \cite{Sikivie:2009qn,Erken:2011dz,Banik:2015ola}. The thermalization process transforms the axions/ALPs system from a coherent system to a thermal system. This "phase transition" can spontaneously break symmetries of the system on large scales \cite{Yang:2015paa}. For example when the thermalization rate of ALPs, $\Gamma\sim Gm^2n_al^2$ where $n_a$ is the local particle density and $l$ is the correlation length, is higher than the Hubble rate the majority of ALPs will be thermalized to form a BEC so most of particles go to the lowest energy state available. This process creates an asymmetry where a preferred momentum is selected. When ALPs are a major component of dark matter, the momentum transit to the baryons and electrons through dynamical friction and the velocity of baryons and electrons was imprinted in the CMB as a dipole asymmetry. As our universe expanding, the velocity of baryons and the frequency of photons were redshifted but the dipole asymmetry remained the same. The dipole asymmetry of the CMB is written as \cite{Gordon:2005ai,Ade:2013nlj}:
\be
\Delta T(\hat n)=(1+A\hat p\cdot \hat n )\Delta T_{iso}(\hat n)
\ee
where the $\hat n$ is the observational direction and the $\hat p$ is the CMB asymmetry dipole direction. Since the size of particle event horizon was about $26{\rm Mpc}$ at recombination and the de Broglie wavelength of the ALPs was smaller than the size of horizon we can constrain the respective mass of ALPs $m\gtrsim 10^{-29}{\rm eV}$ with $A=0.07$.

\section*{Acknowledgments}
We would like to thank Weitian Den and Jianwei Cui for discussions. This work is supported by the Natural Science Foundation of China under grant Number 11305066 and the Fundamental Research Funds for the Central Universities, HUST: No. 2015TS017.

\end{document}